\documentclass
[superscriptaddress,secnumarabic,amssymb,amsmath,nobibnotes,aps,prd,showkeys,showpacs,nofootinbib,onecolumn,eqsecnum]{revtex4}%
\usepackage{graphics}
\usepackage{graphicx}
\usepackage{color}
\usepackage{epsf}
\usepackage{bm}
\usepackage{amsmath,amssymb,amsfonts,mathrsfs,amsthm}
\usepackage{latexsym}
\usepackage{enumerate}
\usepackage{comment}
\usepackage{hyperref}
\usepackage{amsmath}
\usepackage{amsfonts}
\usepackage{amssymb}%
\providecommand{\U}[1]{\protect\rule{.1in}{.1in}}

\newcommand{\be}{\begin{equation}}
\newcommand{\ee}{\end{equation}}
\newcommand{\bea}{\begin{eqnarray}}
\newcommand{\ea}{\end{eqnarray}}
\newcommand{\ben}{\begin{equation*}}
\newcommand{\een}{\end{equation*}}
\newcommand{\bean}{\begin{eqnarray*}}
\newcommand{\eean}{\end{eqnarray*}}
\def\bal#1\eal{\begin{align}#1\end{align}}

\newcommand{\mincir}{\raise
-3.truept\hbox{\rlap{\hbox{$\sim$}}\raise4.truept\hbox{$<$}\ }}
\newcommand{\magcir}{\raise
-3.truept\hbox{\rlap{\hbox{$\sim$}}\raise4.truept\hbox{$>$}\ }}

\begin{document}
\title{Reduced Lagrangians and analytic solutions in Einstein-\ae ther Cosmology}
\author{M. Roumeliotis}
\email{microum@phys.uoa.gr}
\affiliation{Nuclear and Particle Physics section, Physics Department, University of
Athens, 15771 Athens, Greece}
\author{A. Paliathanasis}
\email{anpaliat@phys.uoa.gr}
\affiliation{Instituto de Ciencias F\'{\i}sicas y Matem\'{a}ticas, Universidad Austral de
Chile, Valdivia, Chile}
\affiliation{Institute of Systems Science, Durban University of Technology, POB 1334 Durban
4000, South Africa.}
\author{Petros A. Terzis}
\email{pterzis@phys.uoa.gr}
\affiliation{Nuclear and Particle Physics section, Physics Department, University of
Athens, 15771 Athens, Greece}
\author{T. Christodoulakis}
\email{tchris@phys.uoa.gr}
\affiliation{Nuclear and Particle Physics section, Physics Department, University of
Athens, 15771 Athens, Greece}

\begin{abstract}
We present the solution space of the field equations in the Einstein-\ae ther
theory for the case of a $FLRW$ and a LRS Bianchi Type $III$ space-time. We
also find that there are portions of the initial parameters space for which no
solution is admitted by the reduced equations. The reduced Lagrangians deduced
from the full action are, in general, correctly describing the dynamics
whenever solutions do exist

\end{abstract}
\keywords{Einstein-\ae ther; Cosmology; Lagrangian; analytic solutions}\maketitle
\date{\today}

\section{Introduction}

The main context in modified theories of gravity is that new geometric
invariants are introduced in the Einstein-Hilbert Action of General
Relativity, for a recent review we refer the reader to \cite{mod,bok1}.
Modified theories of gravity have been a subject of special interest for the
scientific community in recent years because they can describe observable
\cite{dataacc1,dataacc2,data1,data2,data3,data4} phenomena with a geometric
approach. More specifically, the geometric invariants which are added to the
Einstein-Hilbert action, are introducing new components in the gravitational
field equations; these change the dynamics of the field equations in such a
way so that the evolution of the latter describes correctly the observed
phenomena \cite{tsuji}.

In this work, we are interested in the Einstein-\ae ther theory
\cite{jacob,DJ,DJ2,Carru,carroll}. In this theory a unitary time-like vector
field, called the \ae ther, is introduced in the gravitational action: This
amendment amounts in adding to the Einstein-Hilbert Action a kinetic term
quadratic in the covariant derivatives of the field as well as a Lagrange
multiplier ensuring that the field is uni-modular. This modification
spontaneously breaks the Lorentz symmetry \cite{ea1}, by selecting a preferred
frame at each point in space-time while preserving local rotational symmetry.
The gravitational field equations are of second-order and correspond to
variations of the action with respect to the metric tensor, the \ae ther field
and the Lagrange multiplier. There are various applications of the
Einstein-\ae ther theory in cosmological studies. It has been used to describe
the later acceleration phase of the universe, as an alternative to dark energy
models \cite{ea2}, and also for the early inflationary epoch
\cite{Barrow:2012qy}. On the other hand, Einstein-\ae ther can be seen as the
classical limit of Ho\v{r}ava gravity \cite{esf,esf1}. Some exact solutions as
well as a qualitative analysis of this theory- either in cosmological studies
on in static spherical symmetric space-times- can be found in
\cite{Sandin:2012gq,Coley:2015qqa,Latta:2016jix,eda,eda2,eda3,eda4,eda5,eda6,eda7,eda8,eda9}
and references therein.

A main concept in physics is the mathematical description of the nature. What
makes the physical science to differ and stand out from the applied
mathematics is the requirement of principles. The principle of stationary
action is the most common and describes well the physical phenomena from all
areas of nature, from Newtonian Mechanics, General Relativity and Quantum
Mechanics. A prerequisite for this principle of stationary action to be
applied is the existence of a Lagrangian function. This function consists of
the Ricci-scalar in the case of Einstein's General Relativity, and is
supplemented by a kinetic term of the \ae ther field and a Lagrange multiplier
ensuring the uni-modular condition for the case of the Einstein-\ae ther theory.

A set of (autonomous) differential equations is not uniquely described by only
one Lagrange function \cite{lan2}. Indeed, it is possible to exist more than
one Lagrange functions; for instance, any nonlinear function of the velocity
$\dot{q}$, i.e. $f\left(  \dot{q}\right)  $, describes the free particle in
Newtonian Mechanics. In General Relativity the Lagrangian density is a
function of the metric tensor and its first and second derivatives ( through
the Ricci-scalar ). This Einstein-Hilbert action principle correctly
reproduces the field equations when all metric components are varied
independently. However, when particular metrics are considered, usually by
imposing some symmetry, the situation is not straightforward: If the symmetry
is imposed to the full action and the redundant coordinates are integrated
out, the reduced action can, of course, be varied only with respect to the
remaining dependent variables. The ensuing equations of motion may or may not
be equivalent to the reduced Einstein equations, i.e the equation resulting
from the imposition of the symmetry to the full Einstein's Field equations.
When they are equivalent, the corresponding reduced Lagrangian is called
valid. A well known example of a non valid case is the class B family of
Bianchi models ( see e.g.\cite{Mac}, \cite{Korf} and references therein). In
this case one can search for another Lagrangian correctly reproducing the
reduced Einstein equations \cite{Chris}. The importance of the existence of a
Lagrangian description of a given set of equations lies in the reach methods
of analytical Mechanics that can be applied in order to study the evolution of
the field equations and their integrability
\cite{int1,int2,int3,int4,int5,int6,int7}. Moreover, in the mini-super-space
approach (where the reduced Lagrangian takes the form of a point Lagrangian),
the quantum analogs of the classical integrals of motion can be used as
supplementary conditions to the Wheeler-DeWitt equations
\cite{min1,min1a,min1b,min2,min3,min4,min5}.

The class of Bianchi spatially homogeneous cosmologies contains many important
cosmological models, including the standard
Friedmann--Lema\^{\i}tre--Robertson--Walker (FLRW) spacetime and the well
known Mixmaster universe (Bianchi IX) \cite{misner1}. In Bianchi models the
spacetime manifold is foliated along the time axis, with three dimensional
homogeneous hypersurfaces \cite{bk1,bk2}. There are nine different homogeneous
hypersurfaces, each invariant under a corresponding three dimensional Killing
algebra \cite{bk1}. The principal advantage of Bianchi cosmological models is
that in these models the gravitational field equations reduce to a system of
coupled, ordinary differential equations with independent variable any
function $f(t)$ parametrizing the \textquotedblleft
time-direction\textquotedblright. Moreover, the Bianchi models are grouped
into two classes, A\ and B (according to the trace $c^{\alpha}_{\alpha\beta}$
of the structure constants tensor being zero or non-zero, respectively); while
each class is divided into several types. As earler mentioned, a main
difference between these two classes is that for the gravitational field
equations which belong to the models of class A the reduced Lagrangian is
valid \cite{macc}, while for the models of class B the corresponding reduced
Lagrangian is, in general, not valid except for cases of further reduction
(LRS space-times, see e.g. \cite{Korf}). However, for the general Type $V$
line-element, although belonging to class B, a valid lagrangian does exist
\cite{Chris}.

In the vacuum and for the Einstein-\ae ther theory we study the existence of
analytical solutions and description of the spacetimes with a reduced
Lagrangian for two spacetimes of special interest, the homogeneous and
isotropic space described by the FLRW line element, and the Bianchi III
spacetime. Both spacetimes have a valid reduced action in the Einstein-Hilbert
case; though the existence of a reduced Lagrangian is not certain in the
Einstein-\ae ther cosmology. The plan of the paper is as follows:

The basic properties and definitions of the Einstein-\ae ther theory are given
in Section \ref{sec2}. We furthermore derive the field equations for the
space-times of our consideration. Sections \ref{sec3} and \ref{sec4} include
the main material of our analysis: we determine the conditions under which
analytic solutions for the field equations in the Einstein-\ae ther theory
exist and present the entire solution space. This is done for the\ FLRW and
the Locally Rotationally Symmetric (LRS) Bianch Typei $III$ space-time,
respectively. Furthermore, we determine the conditions so that the reduced
Lagrangian is valid and, in doing so, we also classify the space-times here
found. The discussion of our results and our conclusions are presented in
Section \ref{sec5}.

\section{Einstein-\ae ther gravity}

\label{sec2}

Let $u^{a}$ be a unitary time-like vector field which describes the \ae ther,
$u^{a}u_{a}=-1$. Thus the Action integral which, uppon variation of the metric
components, gives the gravitational field equations for the Einstein-\ae ther
gravity, is given by the following expression \cite{jacob}%
\begin{equation}
S_{AE}=\int d^{4}x\sqrt{-g}R+\int d^{4}x\sqrt{-g}\left(  K^{\alpha\beta\mu\nu
}u_{\mu;\alpha} u_{\nu;\beta}+\lambda\left(  u^{c}u_{c}+1\right)  \right)  ,
\label{ai.01}%
\end{equation}
where $K^{\alpha\beta\mu\nu}$ describes the coupling between the \ae ther
field and the gravity, defined as%
\begin{equation}
K^{\alpha\beta\mu\nu}\equiv c_{1}g^{\alpha\beta}g^{\mu\nu}+c_{2}g^{\alpha\mu
}g^{\beta\nu}+c_{3}g^{\alpha\nu}g^{\beta\mu}+c_{4}g^{\mu\nu}u^{\alpha}%
u^{\beta}. \label{ai.02}%
\end{equation}
with $c_{1},c_{2},c_{3}$ and $c_{4}$ being dimensionless constants. \ 

The function $\lambda$ in (\ref{ai.01}) is a Lagrange multiplier which ensures
the unitarity of the \ae ther vector field. Finally, $R$ is the Ricci scalar
of the underlying space-time with metric $g_{ab},$ and describes the
Einstein's General Relativity term in the present theory.

The dynamical equations are obtained by demanding stationarity of the Action
integral (\ref{ai.01}) under arbitrary variations with respect to the metric,
$\frac{\delta S_{AE}}{\delta g^{ab}}=0,$ the \ae ther vector field,
$\frac{\delta S_{AE}}{\delta u_{a}}=0$, and the Lagrange multiplier $\lambda$,
$\frac{\delta S_{AE}}{\delta\lambda}=0$. The latter equation provides the
unitary constraint for the \ae ther vector field.

More specifically, variation with respect to the metric tensor gives the
gravitational field equations%
\begin{equation}
{G_{ab}}=T_{ab}^{\ae },\label{ai.03}%
\end{equation}
where $G_{ab}$ is the Einstein tensor and $T_{ab}^{\ae }$ is the \ae ther
energy-momentum tensor defined as
\begin{align}
{T_{ab}^{\ae }} &  =\frac{1}{2}g_{\mu\nu}K^{\alpha\beta\rho\sigma}%
u_{\rho;\alpha}u_{\sigma;\beta}+\frac{1}{2}g_{\mu\nu}\lambda(g^{\alpha\beta
}u_{\alpha}u_{\beta}+1)+\\
&  -c_{1}g^{\alpha\beta}(u_{\alpha;\mu}u_{\beta;\nu}+u_{\mu;\alpha}%
u_{\nu;\beta})-c_{2}g^{\alpha\beta}(u_{\mu;\nu}u_{\alpha;\beta}+u_{\beta
;\alpha}u_{\nu;\mu})-c_{3}g^{\alpha\beta}(u_{\alpha;\mu}u_{\nu;\beta}%
+u_{\mu;\alpha}u_{\beta;\nu})+\nonumber\\
&  -c_{4}u^{\alpha}u^{\beta}u_{\mu;\alpha}u_{\nu;\beta}-c_{4}(g^{\lambda
\alpha}u_{\mu}u^{\beta}u_{\lambda;\nu}u_{\alpha;\beta}+g^{\beta\lambda
}u^{\alpha}u_{\nu}u_{\beta;\alpha}u_{\lambda;\mu})-\lambda u_{\mu}u_{\nu
}+\nonumber\\
&  +(\frac{1}{2}K^{\alpha\beta\lambda\kappa}u^{\rho}u_{\kappa;\beta}g_{\rho
\mu}g_{\lambda\nu}+\frac{1}{2}K^{\alpha\beta\lambda\kappa}u^{\rho}%
u_{\kappa;\beta}g_{\rho\nu}g_{\lambda\mu})_{;\alpha}+(\frac{1}{2}%
K^{\alpha\beta\lambda\kappa}u^{\rho}u_{\kappa;\beta}g_{\rho\mu}g_{\alpha\nu
}+\dfrac{1}{2}K^{\alpha\beta\lambda\kappa}u^{\rho}u_{\kappa;\beta}g_{\rho\nu
}g_{\alpha\mu})_{;\lambda}+\nonumber\\
&  -(\frac{1}{2}K^{\alpha\beta\lambda\kappa}u^{\rho}u_{\kappa;\beta}%
g_{\lambda\mu}g_{\alpha\nu}+\frac{1}{2}K^{\alpha\beta\lambda\kappa}u^{\rho
}u_{\kappa;\beta}g_{\lambda\nu}g_{\alpha\mu})_{;\rho}.\nonumber
\end{align}

Variation with respect to the \ae ther vector field provides the equation of
motion which the vector field $u_{a}$ satisfies, that is,
\begin{equation}
c_{4}g^{\mu\nu}u^{\alpha}u_{\nu;\beta}u_{\mu;\alpha}g^{\kappa\beta}%
-c_{4}g^{\mu\kappa}g^{\alpha\lambda}u_{\lambda;\beta}u^{\beta}u_{\mu;\alpha}-
c_{4}g^{\mu\kappa}u^{\alpha}u^{\beta}_{;\beta}u_{\mu;\alpha} -K^{\alpha
\beta\mu\kappa}u_{\mu;\alpha;\beta}+ \lambda g^{\alpha\kappa}u_{\alpha}=0
\label{ai.04}%
\end{equation}
one component of which can be used to determine the Lagrange multiplier
$\lambda$.

Finally, variation with respect to $\lambda$ gives the condition%
\begin{equation}
{u^{a}}{u_{a}+1=0.} \label{ai.05}%
\end{equation}

It is thus clear that Einstein-\ae ther gravity is a second-order theory; in a
four dimensional manifold this system comprises fifteen equations. It is
important to mention that in our consideration we have not assumed any matter content.

We continue our analysis by selecting the underlying geometry to be that of
(a) FLRW space-time and (b) an LRS Bianchi Type $III$ space-time. In both
cases the field equations are reduced to ordinary, coupled differential
equations with time as the independent (in principle) dynamical variable.

\section{FLRW}

\label{sec3}

The generic line element of the FLRW space-time is taken to be
\begin{equation}
ds^{2}=-M^{2}\left(  t\right)  dt^{2}+a^{2}\left(  t\right)  \left(  \frac
{1}{1-kr^{2}}dr^{2}+r^{2}d\theta^{2}+r^{2}\sin^{2}\theta d\phi^{2}\right)  ,
\label{ai.06}%
\end{equation}
where $a\left(  t\right)  $ denotes the scale factor ( describing the radius
of the three dimensional volume), $M\left(  t\right)  $ is the so-called lapse
function and $k$ characterises the spatial curvature of the three dimensional
hyper-surface ($k=0\Rightarrow$ flat space, $k=1\Rightarrow$ space of constant
positive curvature $k=-1\Rightarrow$ space of constant negative curvature).

We define the \ae ther vector field as $u^{a}=-\dfrac{T^{\prime}\left(
t\right)  }{M\left(  t\right)  ^{2}} \delta_{t}^{a}$, where the prime denotes
derivative with respect to $t$.

$\ $The physical reason for selecting to define the \ae ther vector field with
a derivative is that the curl of $u_{a}$ should be zero so as its potential to
define the physical time. A side effect is that the equations \eqref{ai.03}
become of second order in $t$, a useful occurrence for the reduced Lagrange description.

From the constraint equation (\ref{ai.05}) it straightforwardly follows that%
\begin{equation}
-\left(  \frac{T^{\prime}}{M}\right)  ^{2}+1=0, \label{ai.07}%
\end{equation}
which gives $T\left(  t\right)  =\mu_{1}+\mu_{2}\int M\left(  t\right)  dt$,
$\left(  \mu_{2}\right)  ^{2}=1$. The latter expression quantifies the
preferred frame of the co-moving observer, and in particular the physical time
of the theory $T(t)$.

Furthermore, the only non-vanishing component of (\ref{ai.04}) provides the
Lagrange multiplier $\lambda$ :%
\begin{equation}
\lambda\left(  t\right)  =\frac{3}{a^{2}M^{3}}\left[  \left(  c_{1}%
+c_{2}+c_{3}\right)  Ma^{\prime2}+c_{2}a\left(  a^{\prime}M^{\prime
}-Ma^{\prime\prime}\right)  \right]  . \label{ai.08}%
\end{equation}

Finally, with the use of (\ref{ai.07}) and (\ref{ai.08}), the gravitational
field equations (\ref{ai.03}) become
\begin{equation}
-2kM^{2}+(-2+c_{1}+3c_{2}+c_{3})a^{\prime2}=0,\label{ai.09}%
\end{equation}

\begin{equation}
2kM^{3}+2(-2+c_{1}+3c_{2}+c_{3})aa^{\prime}M^{\prime}-(-2+c_{1}+3c_{2}%
+c_{3})M((a^{\prime})^{2}+2aa^{\prime\prime})=0.\label{ai.10}%
\end{equation}

We now use the freedom to choose the time coordinate and, without loss of
generality, we select for the lapse function $M\left(  t\right)  =a\left(
t\right)  $; this is possible since under an arbitrary change of time $M(t) $,
$a(t) $ transform as density and scalar respectively. Therefore the line
element becomes%
\begin{equation}
ds^{2}=a^{2}\left(  t\right)  \left(  -dt^{2}+\frac{1}{1-kr^{2}}dr^{2}%
+r^{2}d\theta^{2}+r^{2}\sin^{2}\theta d\phi^{2}\right)  , \label{ai.11}%
\end{equation}
and the gravitational field equations reduce to%

\begin{equation}
-2ka^{2}+(-2+c_{1}+3c_{2}+c_{3})a^{\prime2}=0,\label{ai.12}%
\end{equation}

\begin{equation}
(-2+c_{1}+3c_{2}+c_{3})((a^{\prime})^{2}-aa^{\prime\prime})=0.\label{ai.13}%
\end{equation}

\subsection{Analytic solutions}

We proceed by investigating the existence of solutions for the field equations
(\ref{ai.12}), (\ref{ai.13}). The latter implies that our analysis can be
classified according to whether the combination of constants $-2+c_{1}+3
c_{2}+c_{3}$ is equal or different from zero.

\subsubsection{Case 1: $-2+c_{1}+3 c_{2}+c_{3}=0$}

In this case \eqref{ai.13} is identically satisfied, while \eqref{ai.12}
dictates that%
\begin{equation}
k=0,\label{ai.14}%
\end{equation}
thus the solution is described by the above equation and any $a(t)$; this
peculiarity is reflected into the fact that the corresponding reduced
Lagrangian is a total derivative (see next section). The metric becomes
\begin{equation}
\{(-a^{2},0,0,0),(0,a^{2},0,0),(0,0,r^{2}a^{2},0),(0,0,0,r^{2}a^{2}%
sin(\theta)^{2})\},\label{ai.15}%
\end{equation}
and the Lagrange multiplier $\lambda$, is calculated to be%
\begin{equation}
\lambda\left(  t\right)  =-\frac{3((-2+c_{2})(a^{\prime})^{2}+c_{2}%
aa^{\prime\prime})}{a^{4}}.\label{ai.16}%
\end{equation}

\subsubsection{Case 2: $-2+c_{1}+3 c_{2}+c_{3}\neq0$}

1) If k=0 then \eqref{ai.12} implies $a(t)=ca $ which also satisfies
\eqref{ai.13}, while the Lagrange multiplier becomes zero ($\lambda=0$) and
the metric reduces to%

\begin{equation}
\{\{-ca^{2},0,0,0\},\{0,ca^{2},0,0\},\{0,0,ca^{2}r^{2},0\},\{0,0,0,ca^{2}%
r^{2}\sin(\theta)^{2}\}\},\label{ai.17}%
\end{equation}
which of course represents the Minkowski space-time in spherical-polar coordinates.

2) If $\omega\equiv k*(-2 + c_{1} + 3 c_{2} + c_{3})^{-1}>0 $ (the opposite
sign $\omega<0 $ gives an imaginary part to $a(t)$ and is thus not
acceptable), there is a solution $a(t)=m_{1}e^{\pm\sqrt{2} \sqrt{\omega}t} $
It is noteworthy that a linear combination of the above $a(t)$ with different
integration constants $m_{1} , m_{2}$ corresponding to plus and minus signs is
not a solution.

The Lagrange multiplier $\lambda$, is calculated to be%
\begin{equation}
\lambda\left(  t\right)  =\dfrac{6(c_{1}+c_{2}+c_{3})\omega}{a(t)^{2}%
}.\label{ai.18}%
\end{equation}

The metric becomes
\begin{equation}
\{\{-a(t)^{2},0,0,0\},\{0,\dfrac{a(t)^{2}}{1-kr^{2}},0,0\},\{0,0,r^{2}\ast
a(t)^{2},0\},\{0,0,0,a(t)^{2}r^{2}\sin(\theta)^{2}\}\}.\label{ai.19}%
\end{equation}

There is also a solution if $c_{1}+3 c_{2}+c_{3}=0$ which is the Minkowski space-time

\subsection{Reduced Lagrangian description}

For the space-time with line element (\ref{ai.06}) the Ricci-scalar is
calculated as%
\begin{equation}
R=\frac{6a^{\prime\prime}(t)}{a(t)M(t)^{2}}-\frac{6a^{\prime}(t)M^{\prime}%
(t)}{a(t)M(t)^{3}}+\frac{6a^{\prime2}}{a(t)^{2}M(t)^{2}}+\frac{6k}{a(t)^{2}%
}.\label{ai.20}%
\end{equation}

If we substitute in the action (\ref{ai.01}) the above expression,
$u^{a}=-\dfrac{T^{\prime}\left(  t\right)  }{M\left(  t\right)  ^{2}}%
\delta_{t}^{a}$ for the vector field, and ignore a total derivative term we
derive the reduced Lagrangian density%

\begin{align}
\mathcal{L}_{FLRW} &  =\frac{a}{M^{7}}(M^{8}(6k+a^{2}\lambda)-6aM^{5}%
a^{\prime}M^{\prime}-c_{4}a^{2}M^{\prime2}T^{\prime4}+M^{6}(6a^{\prime
2}+a(-a\lambda T^{\prime2}+6a^{\prime\prime}))+\label{ai.21}\\
&  +2c_{4}a^{2}MM^{\prime}T^{\prime3}T^{\prime\prime}-2aM^{3}M^{\prime
}T^{\prime}(3c_{2}a^{\prime}T^{\prime}+(c_{1}+c_{2}+c_{3})aT^{\prime\prime
})+\nonumber\\
&  +a^{2}M^{2}T^{\prime2}((c_{1}+c_{2}+c_{3})M^{\prime2}-c_{4}(T^{\prime
\prime})^{2})\nonumber\\
&  +M^{4}(3(c_{1}+3c_{2}+c_{3})a^{\prime2}T^{\prime2}+6c_{2}aa^{\prime
}T^{\prime}T^{\prime\prime}+(c_{1}+c_{2}+c_{3})a^{2}(T^{\prime\prime}%
)^{2})).\nonumber
\end{align}

In order to test the validity of the reduced Lagrangian density $\mathcal{L}%
_{FLRW}$, we must first derive the corresponding Euler-Lagrange equations.

If these equations are algebraically solved for the accelerations, then
substituted in the reduced equations of the theory \eqref{ai.03},
\eqref{ai.04}, \eqref{ai.05} and the resulting equations are identities we
shall say that $\mathcal{L}_{FLRW}$, is valid since it correctly reproduces
the reduced dynamics.

In the first case $-2+c_{1}+3 c_{2}+c_{3}=0$ after the replacements $T\left(
t\right)  =\mu_{1}+\mu_{2}\int M\left(  t\right)  dt$, $\left(  \mu
_{2}\right)  ^{2}=1$,$M\left(  t\right)  =a\left(  t\right)  $ the Lagrangian
density becomes%

\begin{equation}
\mathcal{L}_{FLRW}=6ka^{2}+6a^{\prime2}+6aa^{\prime\prime}\equiv
6ka^{2}+6(aa^{\prime})^{\prime}.\label{ai.22}%
\end{equation}

It is thus evident that the Euler-Lagrange equation gives $k a = 0$ ,so the
Lagrangian is valid for $k=0 $ and then does not specify a particular $a(t)$,
agreeing with the reduced equations.

In the second case $-2+c_{1}+3 c_{2}+c_{3}\neq0 $ the Lagrangian density
\eqref{ai.21} is always valid.

\section{Bianchi III}

\label{sec4}

We continue our analysis by considering the diagonal LRS Bianchi III spacetime
with fundamental line element%
\begin{equation}
ds^{2}=-M^{2}\left(  t\right)  dt^{2}+a^{2}\left(  t\right)  \left(
dx^{2}+e^{-2x}dy^{2}\right)  +b^{2}\left(  t\right)  dz^{2},\label{ai.23}%
\end{equation}
which admits the following four Killing vector fields:%
\begin{equation}
\xi_{1}=\dfrac{\partial}{\partial y},\quad\xi_{2}=\dfrac{\partial}{\partial
z},\quad\xi_{3}=\dfrac{\partial}{\partial x}+y\dfrac{\partial}{\partial
y},\quad\xi_{4}=2y\dfrac{\partial}{\partial x}+(y^{2}-e^{2x})\dfrac{\partial
}{\partial y}.\label{ai.24}%
\end{equation}
If we apply these symmetries to the \ae ther vector field, demand that the
corresponding one form be curl-free and also utilize \eqref{ai.05} we arrive
at the final form%
\begin{equation}
u^{a}=\dfrac{u_{0}\left(  t\right)  }{-M^{2}}\delta_{t}^{a}+\dfrac{\lambda
_{3}}{b(t)^{2}}\delta_{z}^{a},\qquad u_{0}\left(  t\right)  =\pm M\left(
b\right)  ^{-1}\sqrt{\lambda_{3}^{2}+b^{2}}.\label{ai.25}%
\end{equation}

As we did in the previous case of the FLRW we choose the time so that
$M(t)=a(t)$ which proves equally helpful in the present case as well With this
$u^{a}$ and the line element \eqref{ai.23} the fourth component of the
equation \eqref{ai.04} assumes the form
\begin{equation}
\frac{{\lambda_{3}}\left(  b(t)\lambda(t)-\frac{({c_{1}}+{c_{3}})\left(
a^{\prime}(t)b^{\prime}(t)+a(t)b^{\prime\prime}(t)\right)  }{a(t)^{3}}\right)
}{b(t)^{3}},\label{ai.26}%
\end{equation}
implying that we have to separately investigate the cases $\lambda_{3}=0$,
$\lambda_{3}\neq0$.

The general approach in trying to reveal the solution space is to
algebraically solve two of the equations in terms of the accelerations
$a^{\prime\prime}(t), b^{\prime\prime}(t) $ and substitute the result into the
rest. In doing so some particular branches appear when the denominators of the
corresponding expressions vanish. This may happen either for specific value
ranges of the constants or for particular relations among $a(t) , b(t) $. In
what follows we present all the cases that emerge.

\subsubsection{Case 1: $\lambda_{3}=0, \quad c_{2}=1 $}

The sketch of the solution procedure is as follows:

We solve for $\lambda(t)$ the (0,0) component of \eqref{ai.03} and substitute
into the first component of \eqref{ai.04} which becomes $-\frac{q a^{\prime2}%
}{a(t)^{2}}-\frac{q b^{\prime2}}{2 b(t)^{2}}-1=0 $ where we have replaced
$c_{3}=-c_{1}+q-1 $.

This expression can be satisfied only when $q<0$. In that case the appropriate
scale factors are
\begin{equation}
a(t)=e^{\frac{\int\cos(f(t))\,dt}{\sqrt{-q}}},b(t)=e^{\frac{\sqrt{2}\int
\sin(f(t))\,dt}{\sqrt{-q}},}%
\end{equation}
and the rest of the equations \eqref{ai.03} are satisfied if $f(t)$ obeys the
first order differential equation%

\begin{equation}
f^{\prime}(t)=\frac{\sqrt{2} \cos(f(t))-\sin(f(t))}{\sqrt{-q}}. \label{ai.27}%
\end{equation}

The solution reads%

\begin{equation}
f(t)=-2\tan^{-1}\left(  \frac{1}{6}\left(  3\sqrt{6}\tanh\left(  \frac
{\sqrt{3}\left(  \sqrt{-q}t-m_{1}q\right)  }{2q}\right)  +3\sqrt{2}\right)
\right)  .
\end{equation}

\subsubsection{Case 2: $(\lambda_{3}=0, \quad c_{2}\neq1,\quad\text{and}%
\quad1+c_{1}+c_{3}=0)\quad or\quad( \lambda_{3}=0, \quad c_{2}\neq
1,\quad\text{and}\quad-2+c_{1}+3 c_{2}+c_{3}=0) $}

If we follow analogous steps as before we find that the (3,3) component of
\eqref{ai.03} assumes the never vanishing form $e^{\frac{\sqrt{2} q
t}{(-q)^{3/2}}} $ in the first case and $4 e^{-\frac{2 i t}{\sqrt{-q}}}$ in
the second. Thus there is no solution in these cases.

\subsubsection{Case 3:$\quad\lambda_{3}=0, \quad c_{2}\neq1,\quad
\text{and}\quad(1+c_{1}+c_{3})(-2+c_{1}+3 c_{2}+c_{3})\neq0 $}

If we replace $c_{3} , c_{2} $ as%

\begin{equation}
{c_{3}}=-{c_{1}}+\mu-1~\ ,~~{c_{2}}=\frac{1}{3}\left(  -\frac{\mu\sigma^{2}%
}{2}-\mu+3\right)  ,
\end{equation}
we can solve the non-zero component of \eqref{ai.04} in terms of $\lambda(t)$
and replace into \eqref{ai.03}. As we can easily see the (0,0)component of the
latter equation admits a scaling symmetry $a\rightarrow\omega_{1}a,\quad
b\rightarrow\omega_{2}b$ and does not contain accelerations . Thus, if we make
the replacement $a=e^{\int{a_{1}dt}},b=e^{\int{b_{1}dt}}$ the aforementioned
equation becomes%

\begin{equation}
\frac{1}{12}\left(  4\mu\left(  \sigma^{2}+2\right)  {a_{1}}(t){b_{1}}%
(t)+4\mu\left(  \sigma^{2}-1\right)  {a_{1}}(t)^{2}+\mu\left(  \sigma
^{2}-4\right)  {b_{1}}(t)^{2}-12\right)  =0.\label{ai.28a}%
\end{equation}
which being a quadratic form in $a_{1}(t),b_{1}(t)$ can be solved by%

\begin{equation}
a_{1}(t)=\frac{2 \cosh(f(t))}{\sqrt{3 \mu} \sigma}-\frac{\sigma(2
\sinh(f(t)))}{2 \left(  \sqrt{3 \mu} \sigma\right)  }, b_{1}(t)=\frac{\sigma(2
\sinh(f(t)))}{\sqrt{3 \mu} \sigma}+\frac{2 \cosh(f(t))}{\sqrt{3 \mu} \sigma}
\label{ai.29}%
\end{equation}

If we substitute the above given values of $a_{1}(t), b_{1}(t)$ into the rest
of \eqref{ai.03} we find that a solution exists if $f(t)$ satisfies the
differential equation%

\begin{equation}
f^{\prime}(t)=-\frac{1}{\sqrt{3\mu}\sigma}\left(  \sigma\cosh(f(t))+4\sinh
(f(t))\right)  ,\label{ai.30}%
\end{equation}
with solution
\begin{equation}
f(t)=-2\tanh^{-1}\left(  \frac{4}{\sigma}-\frac{\sqrt{\sigma-4}\sqrt{\sigma
+4}\tan\left(  \frac{1}{6}\left(  3c_{1}\sqrt{\sigma-4}\sqrt{\sigma+4}%
-\frac{\sqrt{3}\sqrt{\sigma-4}\sqrt{\sigma+4}t}{\sqrt{\mu}\sigma}\right)
\right)  }{\sigma}\right)  .\label{ai.31}%
\end{equation}

Thus the final solution is
\begin{equation}
a(\text{t})\text{=}e^{-\frac{6\sigma f(t)}{\sigma^{2}-16}}(\sigma
\cosh(f(t))+4\sinh(f(t)))^{\frac{\sigma^{2}+8}{\sigma^{2}-16}},~b(t)=e^{\frac
{6\sigma f(t)}{\sigma^{2}-16}}(\sigma\cosh(f(t))+4\sinh(f(t)))^{-\frac
{2\left(  \sigma^{2}-4\right)  }{\sigma^{2}-16}},\label{ai.32}%
\end{equation}
where $f(t)$ is to be replaced by the above given value. These cases exhaust
the assumption $\lambda_{3}=0$.

We are thus left to examine the case $\lambda_{3}\neq0$ The strategy is now to
solve the (0,0) component of (\ref{ai.03}) for the Lagrange multiplier
$\lambda(t)$ and substitute into the two components of \eqref{ai.04}. In doing
so some branches appear:

\subsubsection{Case 4: $\lambda_{3}\neq0\quad c_{1}+c_{2}+c_{3}=0\quad c_{2}=0
$}

In this case $\lambda(t)$ reads%

\begin{equation}
\lambda(\text{t})\text{=}\frac{b(t) \left(  -2 a(t) a^{\prime}(t) b^{\prime
}(t)-b(t) a^{\prime2}+a(t)^{2} b(t)\right)  }{a(t)^{4} \left(  b(t)^{2}%
+{\lambda_{3}}^{2}\right)  }. \label{ai.33}%
\end{equation}

When this is substituted into the first and the fourth component of
\eqref{ai.04}, we end up with the single equation.
\begin{equation}
-\frac{2 a^{\prime}(t) b^{\prime}(t)}{a(t) b(t)}-\frac{a^{\prime2}}{a(t)^{2}%
}+1=0. \label{ai.34}%
\end{equation}
As before, there is the scaling symmetry and thus we can make the replacement%

\begin{equation}
a(t)=e^{\int\frac{\sinh(f(t))}{\sqrt{3}}\,dt+\int\frac{\cosh(f(t))}{\sqrt{3}%
}\,dt},b(t)=e^{\int\frac{\cosh(f(t))}{\sqrt{3}}\,dt-2\int\frac{\sinh
(f(t))}{\sqrt{3}}\,dt}\label{ai.35}%
\end{equation}
which satisfies the equation. As far as the equations
\eqref{ai.04},\eqref{ai.03} are concerned, a solution will exist after the
above replacement, if the unique differential equation
\begin{equation}
f^{\prime}(t)=\frac{\cosh(f(t))-2\sinh(f(t))}{\sqrt{3}}\label{ai.36}%
\end{equation}
is satisfied. 

Finally, the solution of \eqref{ai.36} is determined to be%
\begin{equation}
f(t)=2\tanh^{-1}\left(  2-\sqrt{3}\tanh\left(  \frac{1}{2}\left(  \sqrt
{3}c_{1}+t\right)  \right)  \right)  .
\end{equation}

\subsubsection{Case 5: $\lambda_{3}\neq0\quad c_{1}+c_{2}+c_{3}=0\quad c_{2}=1
$}

We now solve the forth component of (\ref{ai.04}) for the Lagrange multiplier
$\lambda(t)$ obtaining
\begin{equation}
\lambda(t)=-\frac{a^{\prime}(t)b^{\prime}(t)+a(t)b^{\prime\prime}(t)}%
{a(t)^{3}b(t)}.\label{ai.37}%
\end{equation}
If we substitute the above $\lambda(t)$ into the first component of
\eqref{ai.04}, we end up with
\begin{equation}
a^{\prime\prime}(t)=\frac{a^{\prime}(t)b^{\prime}(t)}{b(t)}+\frac{a^{\prime2}%
}{a(t)},\label{ai.38}%
\end{equation}
If we substitute the above \eqref{ai.38} into \eqref{ai.03} we observe that
there is no solution since its $(0,0)$ , $(3,3)$ components become the
impossible equations $-\frac{\text{$\lambda$3}^{2}a^{\prime2}}{a(t)^{2}%
b(t)^{2}}-1=0,\quad\frac{\text{$\lambda$3}^{2}a^{\prime2}+a(t)^{2}b(t)^{2}%
}{a(t)^{4}}=0$ respectively.

\subsubsection{Case 6: $\lambda_{3}\neq0\quad c_{1}+c_{2}+c_{3}=0\quad
c_{2}\neq0,c_{2}\neq1 $}

Since $\lambda_{3}\neq0,c_{2}\neq0$, we can solve the fourth component of
(\ref{ai.04}) for the Lagrange multiplier $\lambda(t)$ resulting in
\begin{equation}
\lambda\left(  t\right)  =-\frac{c_{2}\left(  a^{\prime}(t)b^{\prime
}(t)+a(t)b^{\prime\prime}(t)\right)  }{a(t)^{3}b(t)}.\label{ai.39}%
\end{equation}
If we replace (\ref{ai.39}) into the first component of \eqref{ai.04} and
solve it for $a^{\prime\prime}(t)$ we obtain

\begin{center}%
\begin{equation}
a^{\prime\prime}(t)= a^{\prime}(t) \left(  \frac{a^{\prime}(t)}{a(t)}%
+\frac{b^{\prime}(t)}{b(t)}\right)  . \label{ai.40}%
\end{equation}

\end{center}

If $a(t)$ is constant there is no solution since the $(0,0)$ component of
\eqref{ai.03} becomes $-1=0$. For $a(t)$ non-constant, after the replacement
of \eqref{ai.40} into \eqref{ai.03}, its (2,2) component gives%

\begin{equation}
\frac{(c_{2}-1)\left(  a^{\prime}(t)b^{\prime}(t)+a(t)b^{\prime\prime
}(t)\right)  }{a(t)b(t)}=0,\label{ai.41}%
\end{equation}
which is satisfied by the first integral $a(t)b^{\prime}(t)=m$.

If $m=0$ then $b(t)=\text{cb}$ and the replacement to the \eqref{ai.03} make
the (0,0) component
\begin{equation}
\frac{a^{\prime2}\left(  -({c_{2}}-1)cb^{2}-{c_{2}}{\lambda_{3}}^{2}\right)
-cb^{2}a(t)^{2}}{cb^{2}a(t)^{2}}=0.\label{ai.42}%
\end{equation}
This equation is easily integrated resulting in the overall solution
\begin{equation}
a(t)=m_{1}e^{\frac{cbt\epsilon}{\sqrt{-c_{2}cb^{2}-c_{2}\text{$\lambda$}%
_{3}^{2}+cb^{2}}}},b(t)=cb.\label{ai.43}%
\end{equation}

If $m\neq0$ then the integral can be written as $b^{\prime}(t)=\frac{m}%
{a(t)}~$which reduces the first component of \eqref{ai.04} to
\begin{equation}
\frac{2c_{2}\sqrt{b(t)^{2}+\text{$\lambda$}_{3}^{2}}\left(  -a(t)b(t)a^{\prime
\prime}(t)+b(t)a^{\prime2}+ma^{\prime}(t)\right)  }{a(t)^{5}b(t)^{2}%
}=0.\label{ai.44}%
\end{equation}
If the coefficient of $b(t)$ $a^{\prime2}-a(t)a^{\prime\prime}(t)$ vanishes
then the above equation dictates $a(t)=\text{ca}$ and thus $b(t)=\frac{mt}%
{ca}+m_{1}$ which makes the $(0,0)$ component of \eqref{ai.03} $-1=0$
indicating that there is no solution.

Otherwise if $a^{\prime2}-a(t)a^{\prime\prime}(t)\neq0$ then \eqref{ai.44} can
be solved for $b(t)$ giving
\begin{equation}
b(t)=-\frac{ma^{\prime}(t)}{a^{\prime2}-a(t)a^{\prime\prime}(t)}.\label{ai.45}%
\end{equation}
We substitute \eqref{ai.45} into the integral $b^{\prime}(t)=\frac{m}{a(t)}$
and we obtain two valid solutions for $a(t)$.

Finally we use the \eqref{ai.45} obtaining two different solutions with the
final form%

\begin{equation}
a(t)=\frac{e^{-\sqrt{{m_{2}}}({m_{3}}+t)}\left(  2{m_{1}}{m_{2}}%
+e^{\sqrt{{m_{2}}}({m_{3}}+t)}\right)  ^{2}}{4{m_{2}}^{2}},\label{46}%
\end{equation}

\begin{equation}
a(t)=\frac{e^{-\sqrt{{m_{2}}}({m_{3}}+t)}\left(  2{m_{1}}{m_{2}}%
e^{\sqrt{{m_{2}}}({m_{3}}+t)}+1\right)  ^{2}}{4{m_{2}}^{2}},\label{47}%
\end{equation}
where $m_{2}=-\frac{c_{2}{\lambda_{3}}^{2}m_{1}^{2}+m^{2}}{(c_{2}-1)m^{2}}$ in
both the above solutions.

Finally there is the somewhat curious case in which we solve the integral with
respect to $a(t)$ i.e. take $a(t)=\frac{m}{b^{\prime}(t)}$. Then we may solve
algebraically the first component of the \eqref{ai.04} for $b^{(3)}(t)$
\begin{equation}
b^{(3)}(t)=\frac{b^{\prime\prime2}}{b^{\prime}(t)}+\frac{b^{\prime
}(t)b^{\prime\prime}(t)}{b(t)}.\label{ai.48}%
\end{equation}
The apparent branch $b(t)=m_{1}t+m_{2}$ leads to $a(t)=ca$ and has been
earlier seen to lead to no solution.

If we substitute \eqref{ai.48} and $a(t)=\frac{m}{b^{\prime}(t)}$, the (0,0)
component of \eqref{ai.03} gives us
\begin{equation}
-\frac{c_{2}{\lambda_{3}}^{2}b^{\prime\prime2}+b(t)^{2}\left(  (c_{2}%
-1)b^{\prime\prime2}+b^{\prime2}\right)  -2(c_{2}-1)b(t)b^{\prime2}%
b^{\prime\prime}(t)}{b(t)^{2}b^{\prime2}}=0.\label{ai.49}%
\end{equation}
The solution of \eqref{ai.49} as well as $a(t)=\frac{m}{b^{\prime}(t)}$ gives
the final form of the solution
\begin{equation}
a(t)=\frac{2(c_{2}-1)^{2}mm_{1}\cos^{2}\left(  \frac{(m_{2}+t)\sqrt{c_{2}%
}{\lambda_{3}}^{2}+(c_{2}-1)^{2}{m_{1}^{2}}}{2({c_{2}}-1)^{3/2}{m_{1}}%
}\right)  }{{c_{2}}{\lambda_{3}}^{2}+({c_{2}}-1)^{2}{m_{1}}^{2}}.\label{50}%
\end{equation}

We continue with the final case.

\subsubsection{Case 7: $\lambda_{3}\neq0\quad c_{1}+c_{2}+c_{3}\neq0\quad$}

We replace ${c_{3}}\text{=}-{c_{1}}-{c_{2}}+q$ where $q\neq0$. Then the fourth
component of \eqref{ai.04} leads to the following form of the Lagrange
multiplier $\lambda(t)$
\begin{equation}
\lambda\left(  t\right)  =-\frac{({c_{2}}-q)\left(  a^{\prime}(t)b^{\prime
}(t)+a(t)b^{\prime\prime}(t)\right)  }{a(t)^{3}b(t)},\label{ai.51}%
\end{equation}
If we substitute \eqref{ai.51} to the first component of \eqref{ai.04} and to
the (3,3) component of \eqref{ai.03}, a system of equations is created which
can always be solved in terms of $a^{\prime\prime}(t),b^{\prime\prime}(t)$.
\begin{align}
\label{ai.52}
a^{\prime\prime}(t)=\frac{1}{4a(t)b(t)^{2}\left(  b(t)^{2}+{\lambda_{3}}^{2}\right)}
& \left(4{c_{2}}a(t)b(t)a^{\prime}(t)\left(  b(t)^{2}%
+{\lambda_{3}}^{2}\right)  b^{\prime}(t)+2a^{\prime2}\left(  b(t)^{2}%
+{\lambda_{3}}^{2}\right)\left(  b(t)^{2}(c_{2}+q+1)+{\lambda_{3}}^{2}%
(c_{2}+q)\right)  \right. \nonumber \\
& \left. + a(t)^{2}b(t)^{2}\left(qb^{\prime2}+2b(t)^{2}%
+2{\lambda_{3}}^{2}\right)\right)
\end{align}

Replacement of the above values $a^{\prime\prime}(t), b^{\prime\prime}(t)$
into the \eqref{ai.03} we find only the two different equations $(0,0)=0$ and
$(1,1)=0$ quadratic in $a^{\prime}(t), b^{\prime}(t)$. We can solve the
$(0,0)=0$ equation for $b^{\prime2}$ and substitute the result into the first
equation of \eqref{ai.52}, thus obtaining the very simple equation
\begin{equation}
\frac{a^{\prime\prime}(t)}{a(t)}-\frac{a^{\prime}(t) b^{\prime}(t)}{a(t)
b(t)}-\frac{a^{\prime2}}{a(t)^{2}}. \label{ai.53}%
\end{equation}
This equation has the scaling symmetry and can thus be reduced to first order
by the use of the replacement $a(t)=e^{\int{a_{1}}(t) \, dt} $ and ultimately
be solved in terms of $a(t)$ with the result $a(t)=e^{{m_{1}} \int b(t) \,
dt}$

If we substitute this $a(t)$ into (0,0) and (1,1) components of \eqref{ai.03}
and eliminating $b^{\prime2}$ we obtain the following equation
\begin{align}
&  2{m_{1}}\left(  2{c_{2}}^{2}-{c_{2}}(q+4)-q^{2}+q+2\right)  b^{\prime
}(t)+{m_{1}}^{2}b(t)^{2}\left(  2{c_{2}}^{2}-{c_{2}}(q+4)-q^{2}+q+2\right)
+\label{ai.54}\\
&  +2{c_{2}}^{2}{\lambda_{3}}^{2}{m_{1}}^{2}-2{c_{2}}{\lambda_{3}}^{2}{m_{1}%
}^{2}-{c_{2}}{\lambda_{3}}^{2}{m_{1}}^{2}q+2{c_{2}}-{\lambda_{3}}^{2}{m_{1}%
}^{2}q^{2}-q-2=0\nonumber
\end{align}
If ${m_{1}}=0$ then from its definition $a(t)=\text{ca}$ and the above
equation becomes $2{c_{2}}-q-2=0$ which is equivalent to $c_{1}-c_{2}%
+c_{3}=-2$.

The first component of \eqref{ai.04} gives
\begin{equation}
b^{\prime\prime}(t)= \frac{b(t) b^{\prime2}}{b(t)^{2}+{\lambda_{3}}^{2}}
\label{ai.55}%
\end{equation}

We substitute \eqref{ai.55}, ${m_{1}}=0$, $a(t)=\text{ca}$ and $c_{1}%
-c_{2}+c_{3}=-2$ into \eqref{ai.03} and we end up with one differential
equation
\begin{equation}
({c_{2}}-1)b^{\prime2}+b(t)^{2}+{\lambda_{3}}^{2}=0,\label{ai.56}%
\end{equation}
which can be readily solved resulting in the final form of the solution
\begin{equation}
a(t)=ca.\label{56}%
\end{equation}
where $q=2c_{2}-2$. This solution, according to the range of $q$ and the real
or imaginary character of $m_{2}$ is of either neutral or euclidean signature.
If $m_{1}\neq0$ all branches appearing lead to no solutions.

The reduced Lagrangian density is

\begin{center}%
\begin{align}
\mathcal{L}_{III}  &  =\frac{1}{b^{5}M^{7}}(-c_{4}\lambda_{3}^{4}a^{2}%
M^{6}b^{\prime2}-2(c_{1}+c_{3})\lambda_{3}^{2}a^{2}b^{2}M^{6}b^{\prime2}%
+a^{2}b^{4}M^{4}(\lambda_{3}^{2}M^{4}\lambda+(c_{1}+c_{2}+c_{3})u_{0}%
^{2}b^{\prime2})+\label{ai.32}\\
&  +2c_{4}\lambda_{3}^{2}a^{2}b^{3}M^{3}u_{0}b^{\prime}(u_{0}M^{\prime}%
-Mu_{0}^{\prime})+b^{6}(M^{8}(-2+a^{2}\lambda)-4aM^{5}a^{\prime}M^{\prime
}+\nonumber\\
&  -c_{4}a^{2}u_{0}^{4}M^{\prime2}+2c_{4}a^{2}Mu_{0}^{3}M^{\prime}%
u_{0}^{\prime}-2aM^{3}u_{0}M^{\prime}(2c_{2}u_{0}a^{\prime}+(c_{1}+c_{2}%
+c_{3})au_{0}^{\prime})+\nonumber\\
&  +a^{2}M^{2}u_{0}^{2}((c_{1}+c_{2}+c_{3})M^{\prime2}-c_{4}u_{0}^{\prime
2})+M^{4}(2(c_{1}+2c_{2}+c_{3})u_{0}^{2}a^{\prime2}+4c_{2}au_{0}a^{\prime
}u_{0}^{\prime}+\nonumber\\
&  +(c_{1}+c_{2}+c_{3})a^{2}u_{0}^{\prime2})+M^{6}(-a^{2}u_{0}^{2}%
\lambda+2a^{\prime2}+4aa^{\prime\prime}))+\nonumber\\
&  +2ab^{5}M^{3}(-aM^{2}b^{\prime}M^{\prime}-c_{2}au_{0}^{2}b^{\prime
}M^{\prime}+c_{2}Mu_{0}b^{\prime}(2u_{0}a^{\prime}+au_{0}^{\prime}%
)+M^{3}(2a^{\prime}b^{\prime}+ab^{\prime\prime})))\nonumber
\end{align}

\end{center}

The above Lagrangian density can be seen to be valid since the obtained
Euler-Lagrange equations are satisfied by the above given solutions to the
reduced equations .

\section{Conclusions}

\label{sec5} 

In the present work we have investigated the dynamical equations of the
Einstein- \ae ther theory for the case of an $FLRW$ and an LRS Bianchi Type
$III$ geometry. The existence or non-existence of solutions to the reduced
equations depends upon the value of various combinations of the initial
parameters $c_{I},I=1..4$ entering the action integral \eqref{ai.01}:

In the case of $FLRW$ there exist solutions for any value of the parameters,
with the Minkowski metric recovered for the two particular cases $k=0$ and
$c_{1}+3 c_{2}+c_{3}=0$. The case $-2+c_{1}+3 c_{2}+c_{3}=0$ is also
noteworthy since, on the one hand the dynamics forces $k=0$, while on the
other hand $a(t)$ is left free although the time has been fixed by the choice
$M(t)=a(t)$. As for a Lagrangian description $\mathcal{L}_{FLRW}$ correctly
reproduces the reduced dynamics in all cases, even in that where $a(t)$ is
left unspecified.

In the case of LRS Bianchi Type $III$ the situation drastically changes as far
as the existense of solutions to the reduced equations is concerned; there are
considerably large portions of the parameter space for which no solution
exist. For the cases where solutions do exist, the reduced Lagrangian density
$\mathcal{L}_{III}$ correctly recovers them, indicating that it is valid. It
is noteworthy that, when there are no solutions to the reduced equations, The
Lagrangian dynamics given by $\mathcal{L}_{III}$ also leads to no solutions.
Thus, for each of these cases, the corresponding Lagrangian constitutes a
highly non-trivial example of non-compatible dynamics.

As for the physical time inferred from the Einstein-\ae ther one-form $u_{a}$,
it seems that the high degree of symmetry of the $FLRW$ line-element
constrains it to be a function of the time coordinate only $T \equiv T(t)$. In
the case of the LRS Bianchi Type $III$ geometry, the lesser symmetry permits
an initial physical time $T(t,z)\equiv\int u_{0}(t)dt +\lambda_{3} z $; this
shows more resemblance to the Horava gravity.

The pure Einstein-Hilbert solutions are not covered by our method because of
the use of the uni-modularity equation for the connection between $u_{a}(t) $
and the lapse $M(t)$. However, since the Lagrangian densities $\mathcal{L}%
_{FLRW}$ , $\mathcal{L}_{III} $ are valid for all cases in which solutions to
the reduced equations exist, we can easily recover the pure Einstein-Hilbert
solutions by just considering the case $T(t)=\mu_{1} $ for the first and
$u_{0}(t)=0 ,\lambda_{3}=0 $ for the second.

For the future we intend to investigate more Bianchi models and also add extra
matter content e.g. a perfect fluid source in the $FLRW $. It would be
interesting to see which of the properties encountered here persist in these
more general cases.

\begin{acknowledgments}
\noindent AP acknowledges financial support of FONDECYT grant no. 3160121.
\end{acknowledgments}


\end{document}